\documentstyle[preprint,aps,psfig]{revtex}
\begin{document}
\tightenlines
\title{ $\Theta^+$ in a chiral constituent quark model and its
interpolating fields.}
\author{ L. Ya. Glozman}
\address{  Institute for Theoretical
Physics, University of Graz, Universit\"atsplatz 5, A-8010
Graz, Austria\footnote{e-mail: leonid.glozman@uni-graz.at}}
\maketitle


\begin{abstract} 
The recently discovered pentaquark $\Theta^+$ is described
within the chiral constituent quark model. Within this picture
the flavor-spin interaction between valence quarks inverts
the $(1s)^4$ and $(1s)^3(1p)$ levels of the four-quark
subsystem and consequently the lowest-lying pentaquark is
a positive parity , I=0, J=1/2 state of the flavor antidecuplet,
similar to the soliton model prediction. Contrary to the
soliton model, however, the quark picture predicts its spin-orbit
partner with $J=3/2$. Different interpolating fields intended for
lattice calculations of $\Theta^+$ 
are constructed, which have a maximal overlap with this baryon
if it is indeed a quark excitation in the 5Q system.
\end{abstract}
\bigskip

\bigskip
\narrowtext

\section{Introduction}

The recent experimental discovery of the narrow $\Theta^+$ resonance
around 1540 MeV with the strangeness +1 and minimal possible
quark content $uudd\bar s$  \cite{exp1,exp2,exp3,exp4} has 
sharpened the interest in the low
energy QCD spectroscopy and phenomenology. So far its other
quantum numbers
are unknown, except for the isospin, which is probably $I=0$. This
is because this resonance is not seen in the well studied
$I=1$  $K^+ p$ channel.\\

By itself the 5Q component in the baryon wave function is not
something which is very surprising. For example, we know from
the deep inelastic lepton scattering off nucleon that there
is a significant nonstrange antiquark sea component in the
nucleon wave function, implying that on the top of the
valence $QQQ$ component, there are higher $QQQQ\bar Q, ...$
Fock components. Since we know that spontaneous breaking
of chiral symmetry is a key phenomenon to understand the nucleon
and other hadrons in the $u,d,s$ sector in the low-energy
regime, the antiquarks in the nucleon sea are mostly correlated
with quarks to form Goldstone bosons. Consequently the antiquark
polarization in the nucleon sea should not be large \cite{CL}.
This small, but non-zero antiquark polarization can be attributed
to the small amplitude that $Q\bar Q$ are correlated into vector
and higher mesons. The successful description of the low-energy
baryon spectroscopy within the chiral constituent quark model
\cite{GR} also suggests that effects of the higher Fock
components in the baryon wave function are significant. Indeed,
it is a coupling of the valence $QQQ$ component with the 
$QQQ\pi,...$ results in the effective flavor-spin interaction
between the valence quarks, which is attributed to Goldstone
boson exchange \cite{GR}, two-pion-like exchange \cite{RB} or vector
meson-like exchange \cite{G1} between the valence
constituent quarks.
This interaction is known to shift the excited octet of positive
parity (Roper states)
($N(1440)$, $\Lambda(1600)$, $\Sigma(1660),...$) and decuplet states
($\Delta(1600),...$) below the lowest excitations of negative
parity. This physical picture has received  a support 
from recent lattice calculations \cite{DONG,Liu}.\\

Yet, the $\Theta^+$ state is interesting since here the $5Q$
component is a minimal possible Fock component and this state
can belong neither to octet nor decuplet baryons. The minimal
possible representation that can accomodate $S=1$ state is 
antidecuplet.
 If so we can expect
also other antidecuplet members, which, however, can be strongly
mixed with the octet states in those cases where the quantum
numbers of octet and antidecuplet are similar.
The mass and width of the $\Theta^+$ resonance have been
strikingly predicted within the soliton picture \cite{DPP}.\footnote{
The less consistent prediction for the pentaquark mass is given
in ref. \cite{P}.}
Here the main assumption that fixes parameters of the antidecuplet
is that the nonstrange member of the antidecuplet is $N(1710)$.
However, both $N(1710)$ as well as $\Sigma(1880)$ (which is
also considered to be a member of the antidecuplet in \cite{DPP})
are well described within the chiral constituent quark model
as octet states.\\

Both soliton \cite{Sk,Witten} (and quark-soliton \cite{Ripka,B,DP})
as well as chiral constituent quark models \cite{MG,GR,CL2}
rely crucially on spontaneous chiral symmetry breaking as the
most important phenomenon for the baryon physics. Yet, within
the chiral constituent quark model the confinement of quarks is
also considered to be important for radial and orbital
motion of quarks. In the soliton (or quark-soliton)
picture the octet, the decuplet and the antidecuplet members represent
different rotational excitations of the chiral (pion) mean field,
while excited states of positive and negative
parity are considered as resonances in the pion-soliton system
\cite{MK}. The one-particle (quark) motion is not
considered at all within this picture.
Within the chiral
constituent quark model excitations of the nucleon are
either spin-isospin excitations of quarks like in delta,
or radial and orbital excitations of the quark motion
like in $N(1440)$ and in $N(1535)$. That the confining
interaction of quarks should be important for their
orbital motion follows also from the lattice calculations. 
Indeed, at large current quark masses
 excited hadrons can be rigorously described
as a system of quarks with orbital motion in a color-electric 
confining field.
 Lattice calculations show a very smooth evolution
of the $N(1535) - N$ splitting versus current quark masses,
see e.g. \cite{BR}. This splitting is large in the heavy
quark limit and is  described as the orbital excitation
of the quark motion. This splitting very slowly {\it increases} 
towards the chiral limit,
implying that near the chiral limit there is another
mechanism, in addition to confinement, that contributes
to this splitting. This is quite consistent with the
chiral constituent quark model, where it follows that
an appreciable part of this splitting is related to
the flavor-spin interaction between valence quarks \cite{GR,G1}.
Another evidence in favour of the quark picture is that
it provides a remarkably good description of nucleon electromagnetic
and weak formfactors at not very large momenta transfer 
\cite{W}.\\

So it is interesting whether the chiral constituent quark picture
can accomodate $\Theta^+$. It has been realised some time ago
\cite{G2,Stancu} that the lowest pentaquark within this picture
should be of {\it positive} parity, in contrast to pentaquarks
within the naive model (where the residual interaction of constituent
quarks is attributed to perturbative gluon exchange). On the
first value it looks counterintuitive, since naively the ground state
system is expected to be a collection of $1s$ quarks (since the 
intrinsic parity of the antiquark is negative, the ground state
pentaquark must have a negative parity within this picture).
However, if the dominant part of the $N-\Delta$ splitting is
due to a flavor-spin interaction, then this interaction inverts
some of the levels with positive and negative parity (like N(1440)
and N(1535)). The physical reason for such an inversion is
rather simple: The more symmetric the flavor-spin wave function
of the baryon is, the more attractive contribution arises from
the flavor-spin interaction. The Roper state $N(1440)$ and other
similar states belong
to a completely symmetric {\bf 56} representation of $SU(6)$,
while $N(1535)$  and other lowest states of negative parity
are members of
the $SU(6)$ mixed symmetry {\bf 70} plet. Consequently the
flavor-spin interaction shifts the Roper states strongly down
with respect to the negative parity states.\\

Very similar reason explains why the lowest pentaquark is of
positive parity. The orbitally excited pentaquark with
$L=1$ allows for a completely symmetric flavor-spin wave
function of the four-quark subsystem, while such a subsystem
can have only a mixed symmetry if all quarks are in the $1s$
state. Consequently the flavor-spin interaction shifts
the orbitally excited four-quark state below the $(1s)^4$
state. Under some fine tuning of  interaction between
the antiquark and four quarks (which is not constrained by
the usual baryon spectroscopy) it is always possible to provide
the necessary low mass and width of such a pentaquark \cite{SR}.\\

It is interesting that both the soliton picture and chiral
constituent quark picture predict the same quantum numbers
for this antidecuplet state: $I=0,J^P=1/2^+$.
Lattice QCD calculations can potentially answer an important
question about which physical picture is more relevant. Each
picture must imply a very specific interpolator that
optimally creates $\Theta^+$ from the vacuum in the lattice
calculations. It is a purpose of this note to construct the
most optimal interpolators for $\Theta^+$ if this resonance
to be described as a quark (but not soliton) excitatation.

\section{The quantum numbers and wave function of the lowest
pentaquark}

In this section we consider in some detail the lowest
pentaquark state within the chiral constituent quark
model. Consider a $4Q$ subsystem within a pentaquark.
The naive quark model predicts that in the ground state of the
pentaquark all four quarks must be in the same $1s$ state
of orbital  motion and have positive parity. Consequently,
keeping in mind the negative parity of the strange
antiquark, the ground state pentaquark must have negative
parity within the naive model. The orbital wave
function of four quarks must be completely symmetric,
i.e. is described by the $[4]_O$ Young diagram. The color
part of these four quarks has a unique permutational
symmetry $[211]_C$ in order to provide a color-singlet
wave function of the pentaquark

\begin{equation}
[211]_C \times [11]_C = [222]_C + ...
\label{c1}
\end{equation}

\noindent
Hence the combined color-orbital permutational symmetry
of four quarks within the naive model is

\begin{equation}
[211]_C \odot [4]_O = [211]_{CO}.
\label{co1}
\end{equation}

\noindent
In eq. (\ref{c1} ) and below $[k_1,k_2,...]$ (with all $k_i$ being
the non-negative integers $k_1 \geq k_2 \geq k_3 ...$) is a
notation for the Young diagram with $k_1$ boxes in the first
row, $k_2$ boxes  - in the second row, etc,
$\times$ means outer product
of two representations, which is constructed according to
Littlewood's rule, while $ \odot$ denotes inner product of two
representation of the symmetric group, i.e. product of
different wave functions for the same group of particles. The
Pauli principle requires that the total color-orbital-flavor-spin
wave function of four quarks must be antisymmetric, $[1111]_{COFS}$.
This restricts the flavor-spin wave function to be $[31]_{FS}$.\\

Within the chiral constituent quark model the most attractive
contribution from the flavor-spin residual interaction between
valence quarks ,

\begin{equation}
 -\vec \lambda^F_i  \cdot \vec \lambda^F_j 
\vec \sigma_i  \cdot \vec \sigma_j,
\label{GBE}
\end{equation}

\noindent
arises if the flavor-spin Young diagram is completely
symmetric, $[4]_{FS}$. Such a flavor-spin symmetry can be obtained
only if we allow one of the quarks to be in the $1p$ state, i.e.
the orbital momentum of four quarks is $L=1$. Clearly, the kinetic
energy of the $(1s)^3(1p)$ configuration is larger than that of
$(1s)^4$. However, the attraction from the flavor-spin interaction,
which is fixed by $N-\Delta$ splitting, is so strong in the $[4]_{FS}$
case, that it overcomes larger kinetic energy and the four-quark
subsystem with the quantum numbers $L=1,[4]_{FS}$ becomes the ground
state of four quarks. In addition, with the given flavor-spin symmetry
of a few-quark system, the most attractive contribution from the 
interaction (\ref{GBE})  arises when
the flavor permutational symmetry is the most "antisymmetric"
among a few possibilities \cite{GR}.
This uniquely fixes the flavor and spin symmetries of the
ground state four-quark subsystem to be $[22]_F$ and $[22]_S$, respectively.
Since the required pentaquark must have strangeness +1, then the
four-quark subsystem can consist only of $u,d$ quarks and hence
$[22]_F$ symmetry uniquely determines isospin of four quarks to be
$I=0$.
Hence, the quantum numbers of the four-quark subsystem within the
$\Theta^+$ pentaquark are

\begin{equation}
P=-, [211]_C, [31]_O, [1111]_{CO}, [22]_F, [22]_S, [4]_{FS}, L=1, S=0, J=1, I=0.
\label{4Q}
\end{equation}

It is clear from the flavor symmetry of four quarks and their isospin
$I=0$, that the pentaquark must belong to the flavor antidecuplet, because
both flavor antidecuplet and octet are contained in the outer product
of $[22]_F$ (four quarks) and $[11]_F$ (antiquark), but only the antidecuplet
is compatible with the $I=0,S=+1$ quantum numbers.

\section{Interpolating field for the pentaquark}

Now our task is to construct such a local interpolating field
which would have maximal overlap with the wave function (\ref{4Q}).
The four-quark interpolator can be constructed as a product of
two diquark interpolators.\footnote{Here and below under diquark
we understand only a subsystem of two quarks without implying a
diquark clustering.} 
The $P=-,[31]_O$ orbital wave function of four quarks can be obtained
in two different ways. The first way is to construct such a
wave function as a system of two scalar (spatially symmetric, $[2]_O$)
diquarks with $L=1$ relative motion orbital momentum. The corresponding
interpolator then will consist of the product of two 
isoscalar-scalar bilinears (see below). However, such an
interpolator will also couple well to the four-quark
subsystem with two strongly clustered isoscalar-scalar
diquarks. Hence it will be difficult, if impossible, to
distinguishe in lattice calculations with such an interpolator
between the present picture and the picture suggested in ref.
\cite{JW}. The second way to obtain $P=-,[31]_O$  four-quark
wave function is to use
 one diquark which is spatially symmetric, $[2]_O$ (i.e.
it has positive parity), and the other diquark which is spatially
antisymmetric, $[11]_O$, with negative intrinsic 
parity. The corresponding interpolator will strongly couple
to the wave function (\ref{4Q}), but will not couple at all
to the system of two strongly clustered scalar diquarks.
Hence, the strong signal obtained with such an interpolator
would mean that one indeed observes the four-quark subsystem
in the state (\ref{4Q}). Below we will consider in detail
such interpolators.\\

Since the color-orbital wave function of
four quarks is $[1111]_{CO}$, both diquark interpolators must
have  antisymmetric color-orbital structurte, $d \sim [11]_{CO}$.
Hence,
one of the diquarks must be color-antisymmetric, $[11]_C$, and the
other - color-symmetric, $[2]_C$. Both diquarks must be symmetric
in the flavor-spin space, $[2]_{FS}$. This can be provided if each
diquark has the same symmetry in flavor and spin spaces. Both
diquarks must also have equal isospin in order that a total isospin
can be constructed to be 0. Hence there are only two possibilities:

(i)
\begin{equation}
d_1 \sim |P=-, [2]_C, [11]_O, [11]_{CO}, [11]_F, [11]_S, [2]_{FS}, L=1, S=0, 
 J=1, I=0>,
\label{d1}
\end{equation}
\noindent

\begin{equation}
d_2 \sim |P=+, [11]_C, [2]_O, [11]_{CO}, [11]_F, [11]_S, [2]_{FS}, L=0, S=0, 
 J=0, I=0>.
\label{d2}
\end{equation}

\noindent

(ii)
\begin{equation}
d_1 \sim |P=-, [2]_C, [11]_O, [11]_{CO}, [2]_F, [2]_S, [2]_{FS}, L=1, S=1, 
 J=0,1,2,  I=1>,
\label{d11}
\end{equation}
\noindent

\begin{equation}
d_2 \sim |P=+, [11]_C, [2]_O, [11]_{CO}, [2]_F, [2]_S, [2]_{FS}, L=0, S=1, 
 J=1, I=1>.
\label{d22}
\end{equation}

\noindent

Now we will translate the language of orbital, flavor and
spin symmetries into the language of covariant bilinears,
wich is required for lattice calculations.
The local interpolator for the 
$|P=-, [11]_O, [11]_F, [11]_S, L=1, S=0, J=1, I=0\rangle$ diquark
must be isoscalar-vector diquark bilinear field

\begin{equation}
\frac{1}{\sqrt 2} \left[ u^T(x) C \gamma_{\mu} \gamma_5d(x)
- d^T(x) C \gamma_{\mu} \gamma_5u(x) \right].
\label{vec}
\end{equation}

\noindent
Here and below $C$ is charge conjugation matrix and $T$ denotes
transpose of the Dirac spinor. Clearly, for the interpolator one
may use only either one of the terms in (\ref{vec}).\\

The $ |P=+, [2]_O, [11]_F, [11]_S, L=0, S=0, J=0, I=0\rangle$ diquark is
to be interpolated by the isoscalar-scalar bilinear

\begin{equation}
\frac{1}{\sqrt 2} \left[ u^T(x) C  \gamma_5 d(x)
-  d^T(x) C\gamma_5u(x) \right].
\label{sc}
\end{equation}

\noindent
The other possible diquarks,
$| P=-, [11]_O, [2]_F, [2]_S, L=1, S=1, J=0, I=1 \rangle$,

\noindent
$|P=-, [11]_O, [2]_F, [2]_S, L=1, S=1, J=1, I=1 \rangle$ and

\noindent
$|P=-, [11]_O, [2]_F, [2]_S, L=1, S=1, J=2, I=1 \rangle$
can be interpolated by the isovector-pseudoscalar

\begin{equation}
\frac{1}{\sqrt 2} \left[ u^T(x) C   d(x)
+ d^T(x) C u(x) \right]; ~u^T(x) C   u(x) ; ~d^T(x) C   d(x),
\label{ps}
\end{equation}

\noindent
isovector-vector

\begin{equation}
\frac{1}{\sqrt 2} \left[ u^T(x) C \gamma_\mu \gamma_5 d(x)
+ d^T(x) C  \gamma_\mu \gamma_5 u(x) \right]; ~
u^T(x) C \gamma_\mu \gamma_5  u(x) ; ~d^T(x) C  \gamma_\mu \gamma_5  d(x),
\label{isvec}
\end{equation}

\noindent
and isovector-pseudotensor

\begin{equation}
\frac{1}{\sqrt 2} \left[ u^T(x) C \sigma_{\mu \nu} d(x)
+ d^T(x) C  \sigma_{\mu \nu} u(x) \right]; ~
u^T(x) C  \sigma_{\mu \nu} u(x) ; ~d^T(x) C  \sigma_{\mu \nu}  d(x)
\label{ten}
\end{equation}

\noindent
bilinears, respectively.\\

Finally, the $|P=+, [2]_O, [2]_F, [2]_S, L=0, S=1, J=1, I=1 \rangle$ diquark
must be described via isovector-axialvector bilinear field

\begin{equation}
\frac{1}{\sqrt 2} \left[ u^T(x) C \gamma_{\mu} d(x)
+ d^T(x) C  \gamma_{\mu} u(x) \right]; 
u^T(x) C  \gamma_{\mu} u(x) ; d^T(x) C  \gamma_{\mu}  d(x).
\label{av}
\end{equation}

\noindent
Note that each quark field in the bilinears above carries
a color index, which is omitted in this section. Clearly
all these color indices must be contracted into a color-singlet
pentaquark, which will be done in the next section.

\section{The color part of the interpolator}

The next step is to specify color indices of quarks and
to construct a four-quark subsystem with the $[211]_C$
symmetry. This can be done with the help of the Clebsch-Gordan
coefficients of the $SU(3)_C$ group. To specify each representation
(wave function) we will use the following chain of subgroups

$$SU(3)_C \supset O(3)_C \supset O(2)_C.$$

\noindent
Hence the color wave function of one particle (or of a few
particles) is characterised by the permutational
symmetry $[f]_C$ (or by the symbol $(pq)$ which is
uniquely connected to $[f]_C$), by "color orbital momentum"
$L_C$ which specifies representation of $O(3)_C$, and by
its projection $m_C$ that determines representation of
$O(2)_C$. For example, the one-quark field belongs to the
fundamental triplet and is completely specified by $[1]_C,
L_C = 1, m_C = -1,0,1.$ In the following it will be
denoted as $|1m_C\rangle$. The antiquark color-antitriplet
 field is specified by $[11]_C, L_C = 1, m_C = -1,0,1.$
The Clebsch-Gordan coefficient for the $SU(3)_C$ is given
as a product of its scalar factor (which is independent of
index $m_C$) and the Clebsch-Gordan coefficient for $O(3)_C$:

$$\langle [f]_C, L_C, M_C |[f']_C, L'_C,M'_C ; [f'']_C, L''_C,M''_C \rangle
=\langle [f]_C, L_C |[f']_C, L'_C ; [f'']_C, L''_C \rangle
C^{L_CM_C}_{L'_CM'_C ~ L''_CM''_C}.$$

\noindent
Then the color-antisymmetric diquark is constructed as
antisymmetrized product of two quarks

$$d_{CA} \equiv |[11]_C, L_C=1, M_C \rangle =\sum_{M'_C, M''_C}
C^{1M_C}_{1M'_C ~ 1M''_C} |1M'_C\rangle |1M''_C\rangle,$$

\noindent
while the two different color-symmetric diquarks are

$$d'_{CS} \equiv |[2]_C, L_C=0, M_C =0\rangle =\sum_{M'_C, M''_C}
C^{00}_{1M'_C ~ 1M''_C} |1M'_C\rangle |1M''_C\rangle,$$

$$d''_{CS} \equiv |[2]_C, L_C=2, M_C \rangle =\sum_{M'_C, M''_C}
C^{2 M_C}_{1M'_C ~ 1M''_C} |1M'_C\rangle |1M''_C\rangle,$$

\noindent
where $M'_C$ and $M''_C$ are color indices of the first
and second quarks within the given diquark. Then we can construct
the required $|[211]_C, L_C=1, M_C \rangle$ tetraquark out of
two diquarks:

$$
|[211]_C, L_C=1, M_C\rangle = \sqrt{1/6} C^{1 M_C}_{00 ~ 1M_C} 
|[2]_C, L'_C=0, M'_C =0\rangle |[11]_C, L''_C=1, M''_C =M_C \rangle $$

\begin{equation}
+\sqrt{5/6} \sum_{M'_C M''_C} C^{1 M_C}_{2 M'_C ~ 1M''_C} 
|[2]_C, L'_C=2, M'_C \rangle |[11]_C, L_C=1, M''_C \rangle.
\end{equation}

Finally, we have to combine the color wave function of the tetraquark
with the antiquark into a color-singlet pentaquark:

$$ | [222]_C, L^{5Q}_C =0, M^{5Q}_C =0 \rangle =
\sum_{M_C} C^{00}_{1M_C ~ 1-M_C} |[211]_C, L_C=1, M_C\rangle
|[11]_C, L_C=1, -M_C \rangle.$$

A final step is to combine two diquarks according to the possibilities
(i) and (ii) in (\ref{d1}) - (\ref{d22}) and strange antiquark
into a few possible interpolators for a pentaquark. As an
example, we present below one of these interpolators

$$
I_1 = \sum_{M_C,M'_C,M''_C,m'_C,m''_C,t'_C,t''_C  } 
C^{00}_{1M_C ~ 1-M_C} \left\{ \sqrt{1/6}
C^{1 M_C}_{00 ~ 1M_C} C^{00}_{1M'_C ~ 1M''_C} C^{1M_C}_{1m'_C ~ 1m''_C}
\right.$$ 

\begin{equation}
\left.
+ \sqrt{5/6}
C^{1 M_C}_{2  t'_C ~ 1t''_C} C^{2t'_C}_{1M'_C ~ 1M''_C} C^{1t''_C}_{1m'_C ~ 1m''_C} \right \}
\left [u^T_{M'_C}C\gamma_\mu \gamma_5 d_{M''_C} \right ]
\left [u^T_{m'_C}C \gamma_5 d_{m''_C} \right ] \bar s_{-M_C}.
\label{I1}
\end{equation}

\noindent
Other possible interpolator can be obtained, e.g. by
substituting  of the vector diquark in the first square
brackets in eq. (\ref{I1}) by the pseudoscalar one,
$\left [u^T_{M'_C}C u_{M''_C} \right ]$ and of the
scalar diquark in the second brackets - by the axial
vector one, $\left [d^T_{m'_C}C \gamma_\mu d_{m''_C} \right ]$.

\section {Discussion}

We have shown that if the discovered $\Theta^+$ state is
to be described within the chiral constituent quark picture,
then the lowest lying pentaquark must have positive parity,
in contrast with the negative parity of the naive quark model.
Also within our picture the lowest pentaquark will have exactly
the same other quantum numbers as within the soliton picture: 
$I=0, J=1/2$. Contrary to the soliton picture, the quark picture
predicts also its spin-orbit partner with $P=+,J=3/2, I=0, S=1$, 
since the
coupling of $L=1$ tetraquark with the strange antiquark produces
both $J=1/2$ and $J=3/2$ states. Keeping in mind that typically
the spin-orbit splittings in baryon spectroscopy are of the
order of 100 MeV and less, it would be interesting to perform
an experimental search of the $J=3/2$,  $S=+1$ pentaquark in
the region 1400 - 1700 MeV.\\

We have constructed a few interpolating fields intended for
a lattice search of $\Theta^+$, that would have a maximal
overlap with $\Theta^+$ if this state is to be described within
the chiral constituent quark picture. The optimal strategy
would be to use simultaneously a few interpolators and to
calculate a cross-correlation matrix. We anticipate however
a difficulty in these lattice calculations. Usually the signal
from the given state is first detected at rather large
quark masses and then traced towards chiral limit. It is {\it a-priori}
not clear, however, whether the analog of $\Theta^+$ exists
in the heavy quark region. If not, the signal from $\Theta^+$
can appear only below some critical current quark mass.

\acknowledgements
The author is indebted to C.B. Lang for a careful readig of
the manuscript and to D.I. Diakonov for constructive comments. 
The work is supported by the FWF project
P14806-TPH of the Austrian Science Fund.

\end{document}